\documentclass[%
 reprint,
superscriptaddress,
 amsmath,amssymb,
 aps,
]{revtex4-1}
\usepackage{bm}             
\usepackage{graphicx}% Include figure files
\usepackage{dcolumn}%

\begin{document}

% Title of the article
\title{Topologically Protected Doubling of Tilted Dirac Fermions in Two Dimensions}

\author{Tohru Kawarabayashi}
%\thanks{}
\affiliation{Department of Physics, Toho University, Funabashi, 274-8510 Japan}

\author{Hideo Aoki}
%\thanks{}%\altaffiliation{}
\affiliation{Department of Physics, University of Tokyo, Hongo, Tokyo, 113-0033 Japan}
\affiliation{Electronics and Photonics Research Institute, Advanced Industrial Science and Technology (AIST), Tsukuba, Ibaraki, 305-8568 Japan}

\author{Yasuhiro Hatsugai}
%\thanks{}
\affiliation{Department of Physics, University of Tsukuba, Tsukuba, 305-8571 Japan}

% Please select about four verbal keywords for your manuscript.
%\keywords{Dirac fermion, chiral symmetry, fermion doubling, two-dimensional system}

\begin{abstract}
The doubling of massless 
Dirac fermions on two-dimensional lattices is theoretically studied. 
It has been shown that 
the doubling of massless Dirac fermions on a lattice with 
broken chiral symmetry is topologically protected 
even when the Dirac cone is tilted.  This is due to 
the generalized chiral symmetry defined for lattice systems, 
where such models can be generated by a deformation of the 
chiral-symmetric lattice models.  
The present paper shows for two-band lattice models that this is a general 
way to produce systems with the generalized chiral symmetry 
in that such systems can always be transformed back to a lattice model with the conventional chiral symmetry.  
We specifically show that the number of zero modes is an invariant of the transformation, 
leading to the topological protection \`{a} la Nielsen-Ninomiya of the doubling of tilted  and massless
Dirac fermions in two dimensions.
\end{abstract}

\maketitle   % please do not remove

\section{Introduction}

Since the discovery of graphene\cite{NG,Kim}, Dirac electrons in two dimensions have 
attracted much attention in the last decade \cite{Neto,Sarma,Neto2}, which is now extended to 
Weyl and Dirac semimetals in three dimensions \cite{Wan,Young1,Young2,TypeI,TypeII,TypeII-2,Weng,DDSM,AMV}.
In such extensive studies of Dirac fermions, massless Dirac fermions 
have been realized in a wide variety of two-dimensional systems such as an organic compound \cite{Tajima,Kajitaetal,KKSF},
cold atoms on optical lattices\cite{ColdAtom}, and photonic crystals \cite{PC}. There, we can note that
massless Dirac cones can in general be tilted as in the organic metal $\alpha$- (BEDT-TTF)
$_2$I$_3$, where the conventional chiral 
symmetry is broken.
The Dirac cones usually arise in pairs, which is dubbed doubled Dirac 
fermions. The doubling is usually
recognized as a topological consequence of the chiral symmetry \cite{NN,HFA,Hatsugai1,Springer}, so that one might think that the phenomenon would be degraded 
for tilted Dirac cones. 
However, the tilted Dirac fermions do seem to always appear in pairs 
in such materials \cite{Kajitaetal,KKSF}, which raises an interesting 
question of whether the doubling 
remains robust in systems without the conventional chiral symmetry.

To understand the topological stability of the tilted Dirac fermions,
we examine the role of a generalized chiral symmetry introduced 
in Refs.\cite{KHMA,HKA} for the
doubling of Dirac fermions. The generalized chiral symmetry is an extension of the
chiral symmetry, which has been proposed to characterize the tilted Dirac fermions. The
extended symmetry can be defined for lattice models as well as in
continuum models \cite{KAH}. While the previous study \cite{KAH} is rather restricted to the
deformation of lattice models that retains the extended symmetry, here we
examine, from a more general point of view, what can be deduced when 
we require lattice models to preserve 
the extended symmetry. We shall find for two-band models
that, if a lattice model respects the generalized chiral symmetry globally,
the doubling of the tilted Dirac fermions is {\it topologically protected}. 
A key
observation is that the system in such a case can be transformed 
from a lattice
model having the conventional chiral symmetry without changing the number
of fermion species. The present result shows that, even when the conventional chiral
symmetry is broken, the generalized chiral symmetry protects the
topological stability of the doubling of tilted Dirac fermions, which 
would be regarded as a
natural extension of the Nielsen-Ninomiya's theorem \cite{NN} to tilted Dirac fermions in two dimensions.

\section{Generalized Chiral Symmetry}

Let us first introduce an extension of  the conventional chiral symmetry, which we call the 
{\it generalized chiral symmetry}. In general, a system is called chiral symmetric if the Hamiltonian $H$ anti-commutes 
with a chiral operator $\Gamma$ as $\{ H ,\Gamma\} = H\Gamma +\Gamma H = 0$, 
where the chiral 
operator $\Gamma$ is a hermitian operator with $\Gamma^2=1$.  
We have thus 
$$
 \Gamma H \Gamma = -H.
$$
When the conventional chiral symmetry is respected, 
the energy eigenstate $\psi_E$ with the eigenenergy $E$ ($H\psi_E = E\psi_E$) can be related to the eigenstate  
$\psi_{-E}$ having an eigenenergy $-E$ by $\psi_{-E} = \Gamma \psi_E$, since $H\psi_{-E} = H\Gamma \psi_E = -E
\Gamma \psi_E = -E \psi_{-E}$. 
The conventional  chiral symmetry therefore  guarantees  the particle-hole symmetry 
irrespective of the details of the Hamiltonian. 
The conventional chiral symmetry holds for vertical Dirac fermions, typically 
in graphene.  
The effective Dirac field Hamiltonian for a vertical Dirac fermion is generally given by 
$$
 H = v_F\left[(\bm{X}\cdot\bm{\sigma}) \pi_x + (\bm{Y}\cdot \bm{\sigma}) 
\pi_y\right],
$$
where $\bm{\pi} = \bm{p} + e\bm{A}$ with $\bm{A}$ being the vector potential 
is the dynamical momentum, 
$\bm{\sigma} = (\sigma_x, \sigma_y, \sigma_z)$ are 
Pauli matrices, $v_F$ the Fermi velocity, and 
$\bm{X}$ and $\bf{Y}$ are three-dimensional real vectors. 
It is then easy to verify that 
$\Gamma H\Gamma  =-H$ is satisfied with $\Gamma = (\bm{X}\times \bm{Y})\cdot \bm{\sigma}/|\bm{X}\times \bm{Y}|$.

The conventional chiral symmetry can also be defined for lattice models with a bipartite structure, 
such as the honeycomb lattice in two dimensions. For a bipartite lattice having transfer integrals $t_{ab}$ only 
between A and B sub-lattices, the Hamiltonian can be expressed as 
$$
 H=\sum_{a\in A,b\in B} t_{ab} c^\dagger_a c_b + h.c.,
$$
where $c_a(c_b)$ denotes an annihilation operator of an electron on an 
A(B) sub-lattice site.   We can then define a chiral operator as $\Gamma = \exp(i\pi \sum_{b\in B}c_b^\dagger c_b)$ which anti-commutes with the Hamiltonian. 
If we express the Hamiltonian and the chiral operator in a matrix form using a basis ($a_1,\ldots, b_1,\ldots$) with 
$\{a_i\}(\{b_i\})$ the basis on the A(B) sub-lattice in the $i$-th unit cell, we have 
$$ 
H=
 \left(\begin{array}{cc}
 O & T \\
 T^\dagger & O
 \end{array}\right), \quad \Gamma =  \left(\begin{array}{cc}
 I & O \\
 O & -I
 \end{array}\right).
$$
Here $T$ stands for the hopping matrix with $t_{ab}$ between A and B sub-lattices as its elements, and $I$ is the identity matrix. It is then straightforward to see that the equation $H\Gamma + \Gamma H =0$ holds, which 
implies the Hamiltonian $H$
is chiral-symmetric 
irrespective of the details of the matrix elements $t_{ab}$.
The chiral symmetry for 
lattice models has been essential  to the topological protection of the  doubling of fermions in 
two and also in four dimensions \cite{NN,Hatsugai1}. 

\begin{figure}[h]%
\includegraphics[width=\linewidth,%height=\linewidth
]{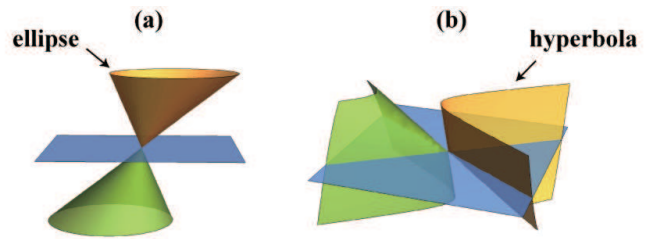}
\caption{%
Schematic figures for the energy dispersions for massless and tilted fermions in the two-dimensional momentum space. 
(a) A tilted Dirac dispersion respecting the generalized chiral symmetry where its cross section with 
a constant-energy plane is an ellipse. Such a plane at the Dirac point 
is represented in blue.   
(b) Energy dispersion for the type-II Dirac fermions for which the cross section 
is given by a hyperbola. The present generalized chiral symmetry is 
applicable only to the elliptic case.
\label{fig0}
}
\end{figure}

For a tilted Dirac fermion, the effective Hamiltonian has additional terms $-v_F(X_0\pi_x + Y_0\pi_y)$ as 
$$
 H =  v_F\left[(-X_0+\bm{X}\cdot\bm{\sigma}) \pi_x + (-Y_0+\bm{Y}\cdot \bm{\sigma}) \pi_y\right].
$$
We have shown that this Hamiltonian satisfies a relation,
$$
 \gamma^\dagger H \gamma  = -H
$$
with $\gamma = T \Gamma T^{-1}$ and $T = \exp(q\bm{\tau}\cdot \bm{\sigma}/2)$ as long as $|\bm{X} \times \bm{Y}|^2- |\bm{\eta}|^2 >0$ 
with $\bm{\eta} \equiv  Y_0\bm{X} - X_0\bm{Y}$ \cite{HKA}. Here 
$\bm{\tau}$ is a unit vector parallel to $(\bm{X}\times\bm{Y})\times\bm{\eta}$, and 
a real parameter $q$ is determined by $\tanh q = |\bm{\eta}|/|\bm{X} \times \bm{Y}|$. This extended symmetry, which we call the 
{\it generalized chiral symmetry}, has been shown to protect the zero-mode Landau levels of tilted Dirac fermions in two dimensions \cite{KHMA}. 
Note that the operator $\gamma$, which we call the {\it generalized chiral operator}, is not hermitian, although we still have $\gamma^2=1$.
The requirement, $|\bm{X} \times \bm{Y}|^2- |\bm{\eta}|^2 >0$, for the generalized chiral symmetry is nothing but 
the geometrical condition that the cross section of the tilted dispersion with a constant energy plane is an 
ellipse (Fig. \ref{fig0}(a)). When  
the dispersion is so much tilted that the cross section becomes an 
open hyperbola (Fig. \ref{fig0}(b)), as in the case of 
the type-II Dirac and Weyl fermions \cite{TypeII,TypeII-2}, the requirement is no longer satisfied, and the present 
generalized chiral symmetry does not apply.

\begin{figure}[h]
\includegraphics[width=\linewidth,%height=\linewidth
]{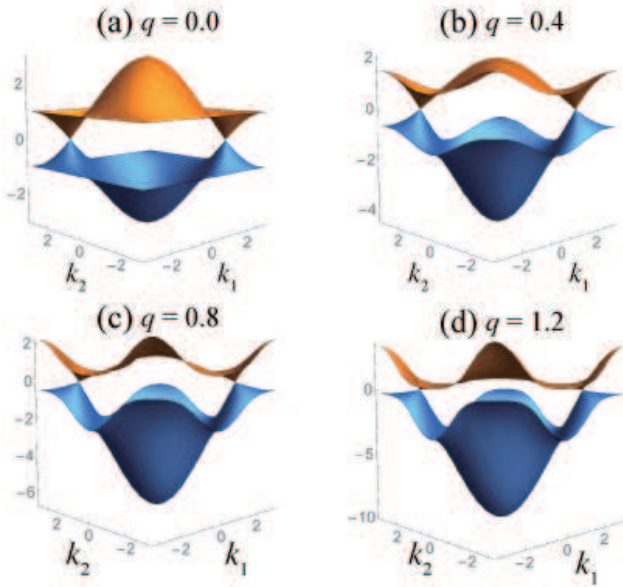}
\caption{%
Energy dispersions are shown for an example series of the lattice models $H=T^{-1}H_cT^{-1}$ transformed from the honeycomb lattice $H_c$ in two dimensions with $T = \exp(q\sigma_x/2)$. The deformation parameter is $q=0.0$ (a), 
$0.4$ (b), $0.8$ (c), and $1.2$ (d), where the Dirac cone is tilted 
for $q\neq 0$. Here the wave vector $k_1(k_2)$ denotes the projection of the wave vector onto the primitive vector $\bm{e}_1(\bm{e}_2)$ of 
the honeycomb lattice \cite{KAH}, and 
the energy in units of the transfer integral between the nearest-neighbor sites of the original $H_c$. The cross section with a constant-energy plane in the vicinity of the contact points remains elliptic 
as long as $q$ is real.
\label{fig_deform}
}
\end{figure}

The generalized chiral symmetry can be defined for lattice models as well. We have shown that a series of lattice models respecting the 
generalized chiral symmetry can be systematically generated by a simple {\it algebraic deformation} of a conventionally 
chiral-symmetric model such as the 
honeycomb lattice model or the $\pi$-flux model on the square lattice \cite{KAH}. Namely, if we have a chiral symmetric 
lattice model with the bipartite structure and denote the Hamiltonian in the momentum space as $H_c$, the algebraic deformation,
$$
 H_n(q) = T^{-1} H_c T^{-1}
$$
with 
$$
 T = \exp(q\bm{n}\cdot \bm{\sigma}/2)
$$
creates a series of deformed Hamiltonian $H_n(q)$. Here $\bm{n}$ is an arbitrary three dimensional real vector with $\bm{n}^2=1$ and 
$q$ a real number. Since the Hamiltonian $H_c$ is chiral symmetric by definition, there exists an operator $\Gamma_c$ satisfying 
$\Gamma_c H_c \Gamma_c = -H_c$ with $\Gamma_c^2 =1$. It is then straightforward to see that we can define a lattice version of the 
generalized chiral operator $\gamma$
as $\gamma = T \Gamma_c T^{-1}$ so that $\gamma^\dagger H_n(q) \gamma = -H$ is satisfied.  
Hence this provides a systematic way for generating lattice models with 
generalized chiral symmetry from those with conventional chiral symmetry \cite{KAH}.
We can note that, when the original chiral-symmetric model has massless Dirac fermions, the deformed lattice models 
respecting the generalized chiral symmetry have still massless but tilted Dirac fermions (Figure \ref{fig_deform}).
We can note that the type-II Dirac fermions do not appear in the present deformation as long as the parameter $q$ is 
real.

However, there remains an important question: Are 
the generalized chiral-symmetric systems generated by the above deformation 
exhaust the possible 
cases that respect the generalized chiral symmetry?  
In other words, it is not obvious that the generalized chiral symmetric lattice model 
can alway be transformed back to a lattice model with the conventional chiral symmetry.  
In the present paper, we precisely consider this question, and we shall 
show for two-band lattice models that a model with the generalized chiral 
symmetry can indeed be always transformed back to a lattice model with the conventional chiral symmetry.  We also show that the 
number of zero modes is an invariant of the transformation, leading to the topological protection of the doubling of tilted  and massless
Dirac fermions in two dimensions.

\section{Generalized Chiral Operator}

For discussing consequences of the generalized chiral symmetry, a general form of the generalized chiral operator
$\gamma$ is crucial. If the system described by the Hamiltonian $H$ is generalized chiral symmetric, 
there exists an operator $\gamma$, not necessarily 
hermitian, that satisfies 
$$
 \gamma^\dagger H \gamma = -H
$$
with $\gamma ^2 = 1$.
Here let us confine ourselves to two-band models where the above equation is described by 2 by 2 matrices in the momentum space.
The general complex 2 by 2 matrix $\gamma$ can be expressed as
$$
 \gamma = \left( \begin{array}{cc}
                   a_{11} & a_{12} \\
                   a_{21} & a_{22}
                   \end{array}\right) = w_0 I_2 +w_x\sigma_x + w_y\sigma_y + w_z\sigma_z ,
$$
where $\{a_{ij}\}\;(i,j = 1,2)$ are complex numbers, $w_0 = (a_{11}+a_{22})/2$, $w_1 = (a_{12}+a_{21})/2$, $w_2 = i(a_{12}- a_{21})/2$, $w_3 = (a_{11}-a_{22})/2$, and 
$I_2$ stands for the 2 by 2 identity matrix.  Here we assume that the elements $\{a_{ij}\}$ for the generalized chiral operator 
$\gamma$ are (complex) constants independent of the momentum.
Since the matrix $\gamma$ must satisfy $\gamma^2=1$, 
we have 
\begin{eqnarray*}
\quad \quad \gamma^2  &=&  (w_0^2+w_x^2+w_y^2+w_z^2)I_2 \\
& &+ 2w_0(w_x\sigma_x+w_y\sigma_y+w_z\sigma_z) \\
&=&I_2,
\end{eqnarray*}
where we have used a relation $\sigma_i \sigma_j = -\sigma_i\sigma_j$ for $i\neq j$.
If $w_0 \neq 0$, then the only solution is trivial 
$\gamma = \pm I_2$, so that we can set $w_0=0$ without loss of generality. The above condition is then reduced to 
$$
 w_x^2+w_y^2+w_z^2 = 1.
$$
If we decompose the complex coefficients $w_i (i=x,y,z)$ 
into real and imaginary parts as
$$
 w_i = w_R^i + i w_I^i,
$$
we end up with two equations,  
$$
 \bm{w}_R^2 - \bm{w}_I^2 = 1
$$
and 
$$ 
 \bm{w}_R \cdot \bm{w}_I = 0,
$$
where $\bm{w}_R = (w_R^x,w_R^y,w_R^z)$ and $\bm{w}_I = (w_I^x,w_I^y,w_I^z)$ are 
three dimensional real vectors. Introducing a real parameter $q$ as 
$$
 \bm{w}_R = (\cosh q) \bm{n}_R, \quad \bm{w}_I = (\sinh q) \bm{n}_I,
$$
we arrive at 
$$
 \bm{n}_R \cdot \bm{n}_I =0
$$
with $\bm{n}_R^2 = \bm{n}_I^2 =1$.
With these parameters, the generalized chiral operator $\gamma$ can be expressed as 
$$
 \gamma = \left[(\cosh q)\bm{n}_R + i(\sinh q)\bm{n}_I\right]\cdot \bm{\sigma} .
$$
This can be rewritten as 
\begin{eqnarray*}
 \gamma & = & (\bm{n}_R\cdot \bm{\sigma})\exp(-q(\bm{\tau}\cdot\bm{\sigma})) \\
  & = & \exp(q(\bm{\tau}\cdot\bm{\sigma})/2)  (\bm{n}_R\cdot \bm{\sigma}) \exp(-q(\bm{\tau}\cdot\bm{\sigma})/2)
\end{eqnarray*}
with 
$$
\bm{\tau} = \bm{n}_R \times \bm{n}_I
$$ because
\begin{eqnarray*}
\lefteqn{(\bm{n}_R\cdot \bm{\sigma})\exp(-q(\bm{\tau}\cdot\bm{\sigma}))} \\
 &=& (\bm{n}_R\cdot \bm{\sigma})\left[\cosh q -\sinh q(\bm{\tau}\cdot\bm{\sigma})\right]\\
  &=&  \cosh q(\bm{n}_R\cdot \bm{\sigma}) -i\sinh q ((\bm{n}_R \times \bm{\tau})\cdot \bm{\sigma})\\
  &=& \left[(\cosh q)\bm{n}_R + i(\sinh q)\bm{n}_I\right]\cdot \bm{\sigma},
\end{eqnarray*}
where we have used the relations $\bm{n}_R \times \bm{\tau} = -\bm{n}_I$ and $(\bm{n}_R\cdot \bm{\sigma})(\bm{\tau}\cdot \bm{\sigma}) = i
(\bm{n}_R\times \bm{\tau})\cdot\bm{\sigma}$. Thus the generalized chiral operator can be written, without loss of generality, as
$$
 \gamma = T_{\bm{\tau}}(q) (\bm{n_R}\cdot\bm{\sigma}) T_{\bm{\tau}}(q)^{-1}
$$
with 
$$
 T_{\bm{\tau}}(q) = \exp(q(\bm{\tau}\cdot\bm{\sigma})/2).
$$

\section{Transformation to Chiral Symmetric Lattice}

With this representation of the generalized chiral operator $\gamma$, 
we can define a deformation from $H$ {\it back to} 
a conventionally chiral-symmetric $H_0$. Indeed, if we define $H_0$ as  
$$
 H_0 = T_{\bm{\tau}}(q) H T_{\bm{\tau}}(q),
$$
it is readily verified that 
$$
 (\bm{n}_R\cdot \bm{\sigma}) H_0 (\bm{n}_R\cdot \bm{\sigma}) = -H_0,
$$
which means that the Hamiltonian $H_0$ respects the conventional chiral symmetry with 
the chiral operator given by $\Gamma = (\bm{n}_R\cdot \bm{\sigma})$.
For two-band models, the generalized chiral symmetric lattice models can thus always be 
transformed back to the chiral symmetric models by an algebraic transformation. 
It is to be noted that the operator $T_{\bm{\tau}}(q)$ satisfies $\det T_{\bm{\tau}}(q) =1$ and the transformation 
back to the conventional chiral symmetry preserves the 
zero-modes of the original Hamiltonian 
$H$.  For instance, if we have an eigenstate $\psi$ with $H\psi = 0$, then $\psi'=T_{\bm{\tau}}(q)^{-1}\psi$ is 
another 
zero-eigenstate of $H_0$ with $H_0\psi'=0$. The number of zero modes is therefore an {\it invariant} of the 
transformation. Since the number of zero modes is equivalent to the number of massless Dirac fermions in  the present case,   
this means that the number of massless Dirac fermions in $H$ has to be the same as that of $H_0$. 

\section{Fermion Doubling}

\begin{figure}[h]%
\includegraphics[width=\linewidth,%height=\linewidth
]{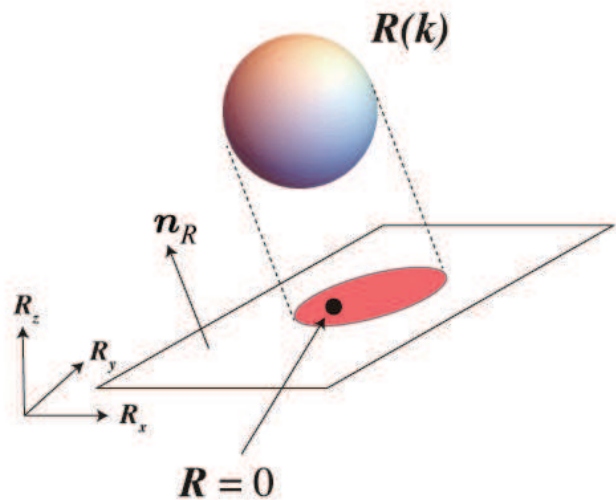}
\caption{%
  Projection of a closed two-dimensional surface $\bm{R}(\bm{k})$ with $\bm{k}$ in the two-dimensional Brillouin 
  zone onto 
  the plane normal to $\bm{n}_R$ in the three-dimensional space of $\bm{R}$ 
is schematically shown. The vector $\bm{n}_R$ is determined by the generalized 
  chiral operator $\gamma$ satisfying $\gamma^\dagger H \gamma = -H$.  If the projected area (red) includes the origin $\bm{R}=0$, we have even numbers 
  of massless Dirac fermions. 
\label{fig_chiral}
}
\end{figure}

Following the argument in Ref. \cite{Hatsugai1}, we can move 
on to the fermion doubling for the chiral symmetric Hamiltonian $H_0$. When the Hamiltonian $H_0$ is expressed as 
$$
 H_0 (\bm{k}) = \bm{R}(\bm{k})\cdot \bm{\sigma}
$$
with $\bm{k}=(k_x,k_y)$ being the two-dimensional 
wave vector, the real three-dimensional coefficient $\bm{R}(\bm{k})$ forms in general a three-dimensional 
surface (closed surface, since the two-dimensional 
Brillouin zone is a closed (torus) space) in the 
space of $(R_x(\bm{k}), R_y(\bm{k}), R_z(\bm{k}))$.  
If $H_0$ 
anti-commutes with a chiral operator $(\bm{n}_R\cdot \bm{\sigma})$, this condition reads 
$$
 \bm{R}(\bm{k}) \cdot \bm{n}_R =0,
$$
which implies that  $\bm{R}(\bm{k})$  is ``flattened" 
onto a plane normal to $\bm{n}_R$ under this condition 
(see Fig.2).  The energy dispersion is given by 
$$
 E(\bm{k}) = \pm \sqrt{R_x(\bm{k})^2 + R_y(\bm{k})^2 + R_z(\bm{k})^2},
$$
so that the contact points $E(\bm{k}_0)=0$ of the massless Dirac fermions is determined by 
$$
\bm{R}(\bm{k}_0) = \bm{0},
$$
namely, the origin of the $\bm{R}$ space. 
Now, for the flattened $\bm{R}$ in the presence of 
the chiral symmetry, for each point $\bm{R}_0$ 
on the flattened object we have always 
an even number of wave vectors satisfying 
$\bm{R}(\bm{k})=\bm{R}_0$.  In particular, if the origin 
is contained in the object, we have an 
even number of wave vectors satisfying $\bm{R}(\bm{k})=0$ (Fig.\ref{fig_chiral}), which is nothing but the doubling of massless Dirac fermions. 
These doubled Dirac fermions are topologically protected, since massless Dirac fermions cannot 
be annihilated as long as the origin is included in the area, while we have no massless Dirac fermions if the origin is outside the area.

Now we discuss the fermion doubling for the Hamiltonian that respects the generalized chiral symmetry.
As shown in the previous section, a generalized chiral-symmetric Hamiltonian $H$ can always be transformed algebraically 
back to the Hamiltonian $H_0$ having the conventional chiral symmetry. Since the number of zero modes is invariant 
of the transformation between $H$ and $H_0 = T_{\bm{\tau}}(q) H T_{\bm{\tau}}(q)$, the number of massless Dirac fermions in $H$
has to be the same as that in $H_0$, which is guaranteed to be an even 
number due to the conventional chiral symmetry. 
Hence we can conclude that any two-band lattice model that respects the generalized chiral symmetry always has doubled tilted Dirac fermions.

\section{Summary}

To understand the stability of the doubled and tilted Dirac fermions as in organic compounds, 
we have investigated the role of the generalized chiral symmetry on the 
fermion doubling in two dimensions, in particular, for the case where the conventional chiral symmetry is broken. 
We have shown for two-band lattice models that the generalized chiral symmetry, defined by the existence of a constant 
2 by 2 matrix $\gamma$ satisfying $\gamma^\dagger H \gamma = -H$ with the Hamiltonian $H$ in the momentun space, protects the topological 
stability of the doubling of the 
tilted and massless Dirac fermions in two dimensions. We have therefore 
relaxed the condition for the topological protection of the fermion doubling from the conventional chiral symmetry to 
the generalized chiral symmetry, which may be thought 
of an extension of Nielsen-Ninomiya's theorem to a certain type of
tilted Dirac fermions in two dimensions.
The present approach to the doubling of Dirac fermions is not applicable to the type-II Dirac fermions 
because the generalized chiral symmetry is no longer available there.

\begin{acknowledgements}
The work was partly supported by  JSPS KAKENHI grant numbers
JP15K05218 (TK), JP16K13845 (YH) and JP17H06138.
\end{acknowledgements}

\end{document}